\documentclass[journal,twoside]{IEEEtran}

\usepackage[dvips]{graphicx,color}
\usepackage{amsfonts}
\usepackage{enumerate}
\newtheorem{thm}{Theorem}
\newtheorem{lem}[thm]{Lemma}

\ifCLASSINFOpdf
\else
\fi
\usepackage[cmex10]{amsmath}
\usepackage{url}

\begin{document}
%
\title{An Improved Analysis of Least Squares Superposition Codes with 
Bernoulli Dictionary}
%
%
%

\author{Yoshinari~Takeishi
        and~Jun'ichi~Takeuchi,~\IEEEmembership{Member,~IEEE,}
\thanks{
This material was presented in part at 
IEEE International Symposium on 
Information Theory 2016, in Barcelona, Spain.
}
\thanks{Y.\ Takeishi
was with Graduate School of Information Science and Electrical Engineering, 
 Kyushu University, Motooka 744, Nishi-ku, Fukuoka-city, 
 Fukuoka 819-0395, Japan. 
(e-mail: takeishi (at) me.inf.kyushu-u.ac.jp).
In April 2013, he moved to 
Mitsubishi Electric Information Network Corporation,
4-6-8 Shibaura Minato-Ku, Tokyo 108-0023, Japan.
}%
\thanks{J.\ Takeuchi 
is with Faculty of Information Science and Electrical Engineering, 
 Kyushu University, Motooka 744, Nishi-ku, Fukuoka-city, 
 Fukuoka 819--0395, Japan. 
(e-mail: tak (at) inf.kyushu-u.ac.jp).}%
}

\maketitle

\begin{abstract}
For the additive white Gaussian noise channel with average power constraint, 
sparse superposition codes, proposed by Barron and Joseph in 2010, achieve the capacity.
While the codewords of the original sparse superposition codes are made with a 
dictionary matrix drawn from a Gaussian distribution, 
we consider the case that it is drawn from a Bernoulli
distribution.
We show an improved  upper bound on its block error probability with 
least squares decoding, 
which is fairly simplified and tighter bound than our previous result in 2014.

\end{abstract}


%

\section{Introduction}
%
%
%
%


\IEEEPARstart{W}{e}
argue the error probability of superposition codes \cite{Barron1,Barron_eff}
with Bernoulli dictionary and least squares decoding.
In this paper, we improve the upper bound of the error probability shown in \cite{IEEE_takeishi}.
The obtained bound is tighter and is in a simpler form than the previous result.

Sparse superposition codes proposed by Barron and Joseph
are applied on the Additive White Gaussian Noise (AWGN) channel 
and shown to achieve the capacity \cite{Barron1,Barron_eff}.
In the coding of sparse superposition codes, 
we generate a real valued matrix, which we call dictionary, 
then make codewords by superposition of column vectors from the dictionary.
Namely, codewords vector $c$ is described 
with the matrix $X$ and a coefficient vector $\beta$ as follows;
\begin{eqnarray*}
c=X\beta.
\end{eqnarray*}

In the original sparse superposition codes, we make a dictionary by drawing from a 
Gaussian distribution. 
Using this {\it Gaussian dictionary}, the error probability with least square decoding 
is shown to be
\begin{eqnarray}\label{JBbound}
O\left(\exp\left\{ - (d (C-R)^2  -\log n/ n) n \right\}\right),  
\end{eqnarray}
where $d$ is a certain positive constant, $n$ is code length, $R$ is a transmission rate, 
and $C$ is a channel capacity \cite{Barron1}.
The bound (\ref{JBbound}) is exponentially small in $n$ 
when $R$ satisfies 
\begin{eqnarray}\label{origbounddash}
|C-R| = \Omega((\log n)^{1/2}/ n^{1/2}).
\end{eqnarray}

However, it is difficult to realize the Gaussian dictionary in a real device 
since the Gaussian random variable can take arbitrary large or small value.
In \cite{IEEE_takeishi}, we studied the case that 
the dictionary is drawn from the unbiased Bernoulli distribution. 
Namely, each entry of the dictionary only takes $+1$ or $-1$ with probability $1/2$, respectively.
We proved that the error probability with Bernoulli dictionary 
with least square decoding is 
\begin{eqnarray}\label{oldbound}
O\left(\exp\left\{ - (d(C-R)^2 -\log n /n^{1/4}) n \right\}\right).
\end{eqnarray}
Although the above bound is worse than (\ref{JBbound}), 
it is exponentially small in $n$ when $R$ satisfies 
\begin{eqnarray}\label{oldbounddash}
|C-R| = \Omega((\log n)^{1/2}/ n^{1/8}).
\end{eqnarray}

To show the above bound, we analyzed the error between binomial and 
Gaussian distributions, where evaluation of sectional measurement 
is one of important factors \cite{IEEE_takeishi}.
However, we found the sectional measurement in the analysis loose.
Concretely we found that it is better to use Euler-Maclaurin formula in that 
analysis.
Then the above bound (\ref{oldbound}) is refined as
\begin{eqnarray}\label{newbound}
O\left(\exp\left\{ - (d(C-R)^2 -1 /n^{1/2}) n \right\}\right).
\end{eqnarray}
Comparing the above bound to (\ref{oldbound}), 
$\log n /n^{1/4}$ is reduced to $1 /n^{1/2}$.
Consequently, the condition (\ref{oldbounddash}) is improved to
\begin{eqnarray}\label{bounddash}
|C-R| =\Omega(1/ n^{1/4}).
\end{eqnarray}

In this paper, we treat the least squares decoder, which is optimal
in terms of  error probability, but computationally intractable.
Efficient decoding algorithms are also researched until now, such as 
\cite{Barbier1, Barbier2, Barron_eff2, Barron_eff3, Barron_eff, Rush, Rush2}.
For the efficient decoding algorithms \cite{Barron_eff}, 
the block error probability is
\begin{eqnarray}\label{effbound}
O\left(\exp\left\{ -d (C_n-R)^2n \right\}\right),
\end{eqnarray}
where $R < C_n < C$ and
\begin{eqnarray}\label{eff_cn}
(C-C_n)/C = O(\log \log n/\log n).
\end{eqnarray}
The above bound is exponentially small 
while there is a considerable gap between $C_n$ and $C$ 
for the practical code length.
It is still an open problem to show that sparse superposition codes
with Bernoulli dictionary achieve the capacity with efficient algorithms.

We review the sparse superposition codes in Section \ref{sec2}. 
In section \ref{sec4}, we show the new upper bound of the error 
probability with Bernoulli dictionary. 
Section \ref{pf_lemmas} provides proofs of 
some lemmas used in Section \ref{sec4}.

\section{Sparse superposition codes}
\label{sec2}
In this section, we review the sparse superposition codes 
and show the performance of Gaussian dictionary
with the least squares estimator.

In the following, `$\log$' denotes the logarithm of base 2 and 
`$\ln$' denotes the natural logarithm. 
Gaussian distribution with mean $\mu$ and variance $\sigma^2$ is denoted by 
$N(\mu, \sigma^2)$.

\subsection{Problem setting}
We consider communication via the AWGN channel. 
Assume that a message is a $K$ bit string $u\in \{0,1\}^K$ and 
that it is generated from the uniform distribution on $\{0,1 \}^K$.
We use a real value vector $c \in \Re^n$ as a codeword to send a 
message. The codeword $c$ is polluted by the Gaussian noise in the channel.
Namely, letting $Y \in \Re^n$ be the output of the channel, we have 
\begin{eqnarray*}
Y=c+\epsilon,
\end{eqnarray*}
where $\epsilon$ is a real number string with length $n$ and each coordinate 
is independently subject to $N(0,\sigma^2)$.
The power of $c$ is defined as $(1/n)\sum_{i=1}^n c_i^2$ and it is constrained to be 
not more than $P$ averagely. We also define a signal-to-noise ratio as $v=P/\sigma^2$.

We consider the task to estimate the message $u$ based on $Y$ and $X$.
Let $\hat{u}$ be an estimated $u$. We call the event $\hat{u}\neq u$ ``block error''.
Further, we define the transmission rate $R$ as $K/n$. 
It is desired that we transmit messages at large $R$ with sufficiently small 
block error probability.
It is well known that at all rate less than
\begin{eqnarray}
C=\frac{1}{2}\log(1+v)\ {\rm (bit/transmission)},\nonumber
\end{eqnarray} 
we can transmit	messages with arbitrary small block error probability 
for sufficiently large $n$.

\subsection{Coding}
\begin{figure}[tb]
\begin{center}
\includegraphics[width=7cm,clip,bb=0 0 1254 763]{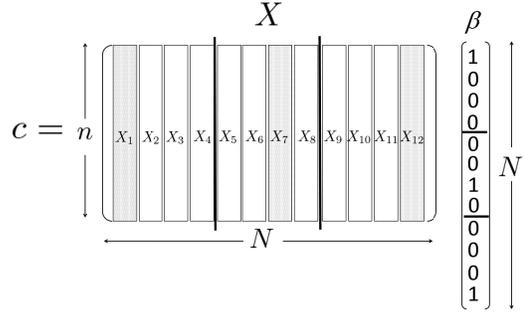}
\caption{Coding method with $L=3$, $M=4$, $N=12$\label{fig1}}
\end{center}
\end{figure}

We state the coding method of sparse superposition codes.
First, we map a message $u$ into a coefficient vector 
$\beta\in \{0,1\}^N$ by a one to one function.
The vector $\beta$ is split into $L$ sections of size $M$ and each section 
has one nonzero element and the other elements are all zero.
Then the codeword $c$ is formed as follows:
\begin{eqnarray}
c=X\beta=\beta_1 X_1+\beta_2 X_2+\cdots+\beta_N X_N,
\nonumber
\end{eqnarray}
where $X$ is an $n\times N$ matrix (dictionary) and 
$X_j$ is the $j$th column vector of $X$.
Thus $c$ is a superposition of $L$ column vectors of $X$, with exactly 
one column selected from each section.
We illustrate an example of coding method in Fig.\ref{fig1}. 


In this paper, we set all nonzero elements 1.
On the other hand, for the efficient decoding algorithms such as 
the addaptive successive decoder proposed in 
\cite{Barron_eff}, 
we set nonzero elements decaying exponentially among sections.
However, we do not treat it here.

In the original paper \cite{Barron1}, each element of the dictionary $X$ is 
independently drawn from $N(0,P/L)$.
This distribution is optimal for the random coding argument 
used to prove the channel coding theorem for the AWGN channel 
with average power constraint by $P$ \cite{Shannon}. 
While in this paper, we analyze the case in which each entry of the dictionary 
is independently drawn as the following random variable:
\begin{eqnarray*}
X_{ij}=
\left\{
\begin{array}{l}
-\sqrt{P/L}\ ({\rm with\ probability}\ 1/2)\\
\sqrt{P/L}\ ({\rm otherwise})
\end{array}
\right.
\end{eqnarray*}
The parameters $L$, $M$, and $N$ are selected so as to satisfy the following.
The number of messages is $2^K$ according to our problem setting about $u$,
and the number of codewords is $M^L$ according to the way of making $\beta$.
Thus we arrange $2^K=M^L$, equivalently, $K = L\log M$. 
According to the original paper \cite{Barron1}, 
the value of $M$ is set to be $L^a$ and the parameter $a$ is referred to 
as section size rate.
Then we have $K=aL\log L$ and $n=(aL\log L)/R$.

\subsection{Decoding}
We analyze the least squares estimator, which makes the error probability 
minimum ignoring computational complexity.
From the received word $Y$ and knowledge of the dictionary $X$, 
we estimates the original message $u$, equivalently, 
estimates the corresponding $\beta$. 

Define a set ${\cal B}$ as
\begin{eqnarray}
{\cal B}=\{ \beta\in \{0,1\}^N| \beta_j\ {\rm has\ one\ }1\ {\rm in\ each\ section} \}.
\nonumber
\end{eqnarray}
Then the least squares decoder $\hat{\beta}$ is denoted as 
\begin{eqnarray}
\hat{\beta}=\arg \min_{\beta\in {\cal B}}\| Y-X\beta\|^2,
\nonumber
\end{eqnarray}
where $\| \cdot \|$ denotes the Euclidean norm. 


Let $\beta^*$ denote the true $\beta$, 
then the event $\hat{\beta}\neq \beta^* $ corresponds to the block error. 
Let $mistakes$ denote the number of sections in which 
the position of the nonzero element in $\hat{\beta}$ is different from 
that in the $\beta^*$. 
Define the error event
$
{\cal E}_{\alpha_0}=\{mistakes\geq \alpha_0 L\},
$ that the decoder makes $mistakes$ in at least $\alpha_0$ fraction of 
sections. 
A proportion of $mistakes$ $\alpha=mistakes/L$ is called section error rate.

\subsection{Performance}
It is proved in the paper \cite{Barron1} 
that given $0<\alpha_0\leq 1$, 
the probability of the event 
${\cal E}_{\alpha_0}$
is exponentially small in $n$. 
The following theorem 
(Proposition~1 in \cite{Barron1})
provides an upper bound on 
the probability of the event ${\cal E}_{\alpha_0}$, 
where 
\begin{eqnarray}
w_v=\frac{v}{[4(1+v)^2]\sqrt{1+(1/4)v^3/(1+v)}}
\nonumber
\end{eqnarray}
and $g(x)=\sqrt{1+4x^2}-1$. 
It follows that 
\begin{eqnarray}
g(x)\geq \min\{\sqrt{2}x,x^2\}\ \ {\rm for\ all }\ \ x\geq 0.
\nonumber
\end{eqnarray}
The definition of $a_{v,L}$ in the statement 
is given later as (\ref{avl}). 

\begin{thm}[Joseph and Barron 2012]
\label{thm1}
Suppose that each entry of $X$ is independently 
drawn from $N(0,P/L)$. 
Assume $M=L^a$, where $a\geq a_{v,L}$, and the 
rate $R$ is less than the capacity $C$. 
Then
\begin{eqnarray}
\Pr[{\cal E}_{\alpha_0}]=e^{-nE(\alpha_0,R)}
\nonumber
\end{eqnarray}
with $E(\alpha_0,R)\geq h(\alpha_0,C-R)-(\ln (2L))/n$, 
where 
\begin{eqnarray}
h(\alpha,\Delta)=\min\left\{ 
\alpha w_v \Delta,\frac{1}{4}g\left(\frac{\Delta}{2\sqrt{v}}\right) \right\}
\nonumber
\end{eqnarray}
is evaluated at $\alpha=\alpha_0$ and $\Delta=C-R$.
\end{thm}
\stepcounter{pf}
{\it Remark}: In this theorem, the unit of $R$ and $C$ is nat/transmission. 
Then, since $n=(aL\ln L)/R$, $L$ is bounded by 
$nR/a$ when $\ln L\geq 1$.

As noted in Joseph and Barron \cite{Barron1}, in order to bound the block error probability, 
we can use composition with an outer Reed-Solomon (RS) 
code \cite{RS} of rate near one. 
If $R_{outer}=1-\delta$ is the rate of an RS code, with $0<\delta<1$, 
then section error rates less than $\delta/2$ can be corrected.
Thus, 
through concatenation with an outer RS code, 
we get a code with rate $(1-2\alpha_0)R$ and block error probability
less than or equal to $\Pr[{\cal E}_{\alpha_0}]$.
Arrange as $R=C-\Delta$ and $\alpha_0=\Delta$, with $\Delta>0$. 
Then the overall rate $(1-2\Delta)(C-\Delta)$ continues to have drop 
from capacity of order $\Delta$. The composite code have block error 
probability of order 
$\exp\{-nd\Delta^2\}$, 
where $d$ is a positive constant.

To prove Theorem \ref{thm1}, 
we evaluate the probability of the event 
$E_l=\{mistakes=l\}$ for $l=1,2,\ldots,L$. 
The probability $\Pr[E_l]$ is used to evaluate 
\begin{eqnarray}
\Pr[{\cal E}_{\alpha_0}]=\sum_{l\geq \alpha_0 L}\Pr[E_l].
\nonumber
\end{eqnarray}


We introduce the function $C_\alpha=(1/2)\ln(1+\alpha v)$ 
for $0\leq\alpha\leq 1$. 
It equals the channel capacity $C$ when $\alpha=1$.
Then $C_{\alpha}-\alpha C$ is a nonnegative function 
which equals $0$ when $\alpha$ is $0$ or $1$ and is strictly positive in between. 
Thus the quantity $C_{\alpha}-\alpha R$ is larger than $\alpha(C-R)$, 
which is positive when $R<C$.

For a positive $\Delta$ and $\rho\in[-1,1]$, 
we define a quantity $D(\Delta,1-\rho^2)$ as
\begin{eqnarray}
D(\Delta,1-\rho^2)=\max_{\lambda\geq 0}
\Bigl\{ \lambda\Delta + \frac{1}{2}\ln(1-\lambda^2(1-\rho^2)) \Bigr\}
\label{eq:D}
\end{eqnarray}
and $D_1(\Delta,1-\rho^2)$ as
\begin{eqnarray}
D_1(\Delta,1-\rho^2)=\max_{0\leq\lambda\leq 1}
\Bigl\{ \lambda\Delta + \frac{1}{2}\ln(1-\lambda^2(1-\rho^2)) \Bigr\}.
\label{eq:D1}
\end{eqnarray}
Note that these quantities are nonnegative. 

The following lemma (Lemma~4 in \cite{Barron1}) 
provides an upper bound on $\Pr[E_l]$. 
\begin{lem}[Joseph and Barron 2012]
\label{lem3}
Suppose that each entry of $X$ is independently 
drawn from $N(0,P/L)$. 
Let a positive integer $l\leq L$ be given and let $\alpha=l/L$. 
Then, $\Pr[E_l]$ is bounded by the minimum for $t_{\alpha}$ in the 
interval $[0,C_{\alpha}-\alpha R]$ of $err_{Gauss}(\alpha)$, where 
\begin{eqnarray}
err_{Gauss}(\alpha)&=&{_LC_{\alpha L}}
\exp\{ -nD_1(\Delta_{\alpha},1-\rho_1^2) \}\nonumber\\
&&+\exp\{ -nD(t_{\alpha},1-\rho_2^2) \}\label{lem2equ}
\end{eqnarray}
with $\Delta_{\alpha}=C_{\alpha}-\alpha R-t_{\alpha}$, 
$1-\rho_1^2=\alpha(1-\alpha)v/(1+\alpha v)$, and 
$1-\rho_2^2=\alpha^2 v/(1+\alpha^2 v)$.
\end{lem}
\stepcounter{pf}

To make (\ref{lem2equ}) exponentially small, it is sufficient that 
the section size rate $a$ is larger than
\begin{eqnarray}
a_{v,L}=\max_{\alpha \in \{\frac{1}{L},\frac{2}{L},\ldots, 1-\frac{1}{L}\}}
\frac{R\ln {}_LC_{L\alpha}}{D_1(C_{\alpha}-\alpha C,1-\rho_1^2)L\ln L}.
\label{avl}
\end{eqnarray} 
The quantity $a_{v,L}$ converges to a finite value as $L$ goes to infinity
(see Lemma 5 in \cite{Barron1}). 

\section{Main results}
\label{sec4}
In this section, we analyze the performance of sparse superposition codes 
with Bernoulli dictionary. 
The result stated here is an improvement of the result in 
\cite{IEEE_takeishi}, where we use the same code.
We improve the upper bound of the error probability by refining 
some lemmas used in \cite{IEEE_takeishi}. 
First, we state the main theorem in this paper.

\begin{thm}
\label{thm2}
Suppose that each entry of $X$ is independent equiprobable $\pm\sqrt{P/L}$. 
Assume $M=L^a$, where $a\geq a_{v,L}$, and rate $R$ is less than capacity $C$. Then,
\begin{eqnarray}
\Pr[{\cal E}_{\alpha_0}]=e^{-nE(\alpha_0,R)}
\nonumber
\end{eqnarray}
with 
\begin{eqnarray}
E(\alpha_0,R)\geq h(\alpha_0,C-R)-(\ln (2L))/n
-\iota (L),
\nonumber
\end{eqnarray}
where $\iota(L)=\max\{\iota_1,\iota_2\}$, which are defined in Lemma \ref{lem5}. 
\label{thm2}
\end{thm}
\stepcounter{pf}

{\it Remark:} 
This theorem is the correspondent of Theorem \ref{thm1} in Bernoulli dictionary case
and the error exponent is worse than that in Theorem \ref{thm1} by $\iota(L)$.
This theorem is the same form as the previous result in Theorem 5 in \cite{IEEE_takeishi},
however $\iota(L)$ converges to zero more rapidly than that in 
the previous result as details mentioned later. 

In order to prove Theorem \ref{thm2}, we use the following lemma, which 
is the correspondent in this case to Lemma \ref{lem3}. 
The definition of $\iota_1$ and $\iota_2$ in Theorem \ref{thm2} 
is given in the following lemma. 
\begin{lem}
\label{lem5}
Suppose that each entry of $X$ is independently equiprobable $\pm\sqrt{P/L}$. 
Let $\alpha_0$ be a certain real number in $(0,1]$
and $\alpha = l/L$. 
Then, for every $L \geq 2$ and 
for all $l$ such that $\alpha_0 \leq \alpha \leq 1$,
$\Pr[E_l]$ is bounded by the minimum 
for $t_{\alpha}$ in the interval $[0,C_{\alpha}-\alpha R]$ 
of $err_{Ber}(\alpha)$, where 
\begin{eqnarray}
err_{Ber}(\alpha)&=&{_LC_{\alpha L}}
\exp\{ -n(D_1(\Delta_{\alpha},1-\rho_1^2)-\iota_1) \}\nonumber\\
&&+\exp\{ -n(D(t_{\alpha},1-\rho_2^2)-\iota_2) \}
\nonumber
\end{eqnarray}
with $\Delta_{\alpha}=C_{\alpha}-\alpha R-t_{\alpha}$, 
$1-\rho_1^2=\alpha(1-\alpha)v/(1+\alpha v)$, and 
$1-\rho_2^2=\alpha^2 v/(1+\alpha^2 v)$. 
The variables $\iota_1$ and $\iota_2$ are defined by 
the following series of equations
\begin{eqnarray*}
\iota_1&=&\ln ((1+\iota_3)(1+\max\{\iota_4,\iota_5\}))\\
\iota_2&=&\phi(L)+\ln 
\left(1+\frac{2\eta}{L} \right)
\end{eqnarray*}
where 
\begin{eqnarray*}
1+\iota_3&=&\max_{\alpha_0 L\leq l \leq L}
\left(e^{\phi(l)}\left(1+\frac{\eta (1+v)}{l}\right)\right) \\
1+\iota_4&=&\!\!\!\!\!\!\!
\max_{\alpha_0 L\leq l \leq L-\sqrt{L}}
\left(e^{\phi(l)+\phi(L-l)}\!\left(1+\frac{\eta}{l}\right)\!\!
\left(1+\frac{\eta}{L-l}\right)\right) \\
1+\iota_5&=&\max_{ L-\sqrt{L}\leq l \leq L-1}
\left(\frac{e^{\phi(l)}}{\sqrt{1-1/\sqrt{L}}}
\left(1+\frac{\eta}{l}\right)\right),
\end{eqnarray*}
$\eta = \sqrt{9/(8\pi e)}$, 
and the function $\phi$ is defined in Lemma 
\ref{lem_prop}.
\end{lem}

{\it Remark:}
The function $\phi$ is $O(1/L)$ by Lemma \ref{lem_prop}.
Thus we have $\iota_1=O(1/\sqrt{L})$ and  $\iota_2=O(1/L)$.
So $\iota=\iota(L)$ in Theorem \ref{thm2} is $O(1/\sqrt{L})$.
In the previous result in \cite{IEEE_takeishi}, 
the order of $\iota$ was $O(\sqrt{\ln L}/L^{1/4})$.
Thus $\iota$ in this paper goes to $0$ faster than that in the previous paper.

To prove this lemma, we evaluate the difference between binomial distribution and 
Gaussian distribution.
We do it by the following two steps. 
The first step is evaluating the proportion of the probability mass function of 
binomial distribution to the probability density function of Gaussian.
The following lemma is given in \cite{IEEE_takeishi} to evaluate that.

\begin{lem}[Takeishi et.al 2014]
\label{lem_prop}
For any natural number $l$, 
\begin{eqnarray*}
\max_{k\in\{0,1,\ldots,l\}}\frac{{}_lC_k (1/2)^l}{N(k| l/2,l/4)}
\leq \exp\{\phi(l)\}
\end{eqnarray*}
holds, where 
\begin{eqnarray*}
\phi(l)=\inf_{\zeta\in(0,1/2)}\phi_{\zeta}(l),
\end{eqnarray*}
\begin{eqnarray*}
\phi_{\zeta}(l)=\max\left\{ \left(\frac{3}{16}c_{\zeta}^2
+\frac{1}{12}\right)\frac{1}{l},\ 
 -\frac{4\zeta^4}{3}l+\ln\frac{l}{2}+\frac{1}{12l},\right.\\ 
\left. -\left(\ln 2-\frac{1}{2}\right)l+\frac{1}{2}\ln\frac{\pi l}{2}
 \right\}
\end{eqnarray*}
and $c_{\zeta}=1/(1+2\zeta)^2+1/(1-2\zeta)^2$. 
In particular, 
for any $l \geq 1000$, it follows that  
$
\phi(l) \leq 5/l.
$
\end{lem}
\stepcounter{thm}

The second step is to evaluate the error in replacing summation about 
discrete random variable with integral about continuous random variable.
It is a feasible way to replace the summation with the integral by the 
sectional measurement.
In the previous result \cite{IEEE_takeishi}, 
they evaluated the error in the sectional measurement
by Lemmas 8, 9, and 10 in \cite{IEEE_takeishi}.
In this paper, we improve these lemmas.
The following lemma is an improvement of Lemma 8 in \cite{IEEE_takeishi}.

\begin{lem}
\label{lem_sm}
For a natural number $n$, let $h=2/\sqrt{n}$ and 
$x_k=h(k-n/2)$ ($k=0,1,\ldots,n$). 
For $\mu\in\Re$ and $s>0$, define 
\begin{eqnarray*}
I_d&=&h\sum_{k=0}^n \exp\left\{-\frac{s^2}{2}(x_k-\mu)^2\right\},\\
I_c&=&\int_{-\infty}^{\infty}
\exp\left\{-\frac{s^2}{2}(x-\mu)^2\right\}dx.
\end{eqnarray*}
Then, we have 
\begin{eqnarray*}
I_d\leq \left(1+\frac{\eta s^2}{n}\right)I_c, 
\end{eqnarray*}
where $\eta=\sqrt{9/(8\pi e)}\leq 0.37$.
\end{lem}
\stepcounter{pf}

Further, by reconsidering the proof and using Lemma \ref{lem_sm}, we also improve Lemmas 9 and 10 
in \cite{IEEE_takeishi}.
The following lemmas are improvements of Lemmas 9 and 10 in \cite{IEEE_takeishi}, respectively.

\begin{lem}
\label{lem_sm2}
For a natural number $n$, define 
$h=2/\sqrt{n}$ and ${\cal X}=\{h(k-n/2)\mid k=0,1,\ldots,n\}$. 
Further, for a 2-dimensional real vector ${\bf x}=(x_1,x_2)^T$ 
and a strictly positive definite $2\times 2$ matrix $A$, define 
\begin{eqnarray*}
I_d=\int_{-\infty}^{\infty}h\sum_{x_1\in {\cal X}}
\exp\left\{-\frac{{\bf x}^TA{\bf x}}{2}\right\}dx_2
\end{eqnarray*}
and 
\begin{eqnarray*}
I_c=\int_{-\infty}^{\infty}\int_{-\infty}^{\infty}
\exp\left\{-\frac{{\bf x}^TA{\bf x}}{2}\right\}dx_1dx_2.
\end{eqnarray*}
Then, we have
\begin{eqnarray*}
I_d\leq \left(1+\frac{\eta A_{11}}{n}\right)I_c, 
\end{eqnarray*}
where $\eta=\sqrt{9/(8\pi e)}\leq 0.37$ and $A_{11}$ is (1,1) element of 
matrix $A$.
\end{lem}
\stepcounter{pf}

\begin{lem}
\label{lem_sm3}
For natural numbers $n$ and $n'$, define 
${\cal X}_1=\{h_1(k-n/2)\mid k=0,1,\ldots,n\}$ and 
${\cal X}_2=\{h_2(k-n'/2)\mid k=0,1,\ldots,n'\}$, 
where $h_1=2/\sqrt{n}$ and $h_2=2/\sqrt{n'}$. 
Further, for a 3-dimensional real vector ${\bf x}=(x_1,x_2,x_3)^T$ 
and a strictly positive definite $3\times 3$ matrix $A$, define 
\begin{eqnarray*}
I_d=\int_{-\infty}^{\infty}
h_1h_2\sum_{x_1\in {\cal X}_1}\sum_{x_2\in {\cal X}_2}
\exp\left\{-\frac{{\bf x}^TA{\bf x}}{2}\right\}dx_3
\end{eqnarray*}
and 
\begin{eqnarray*}
I_c=\int_{\Re^3}
\exp\left\{-\frac{{\bf x}^TA{\bf x}}{2}\right\}d{\bf x}.
\end{eqnarray*}
Then, we have
\begin{eqnarray*}
I_d\leq \left(1+\frac{\eta A_{11}}{n}\right)\left(1+\frac{\eta A_{22}}{n'}\right)I_c, 
\end{eqnarray*}
where $\eta=\sqrt{9/(8\pi e)}\leq 0.37$ and $A_{ij}$ is $(i,j)$ element of matrix $A$.
\end{lem}
\stepcounter{pf}

\subsection{Proof of Lemma \ref{lem5}}
We prove Lemma \ref{lem5} along the lines of the proof of 
Lemma 6 in \cite{IEEE_takeishi}, which is based on Lemma 4 in \cite{Barron1}.

We evaluate the probability of the event $E_l$. 
The random variables are the dictionary $X=(X_1,X_2,\ldots,X_N)$ and 
the noise $\epsilon$. 

For $\beta\in {\cal B}$, let $S(\beta)=\{j| \beta_j=1\}$ denote the set 
of indices $j$ for which $\beta_j$ is nonzero. Further, let 
$
{\cal A}=\{S(\beta)| \beta\in {\cal B}\}
$ 
denote the set of allowed subsets of terms. 
Let $\beta^*$ denote $\beta$ which is sent, and let $S^*=S(\beta^*)$. 
Furthermore, for $S\in {\cal A}$, let $X_S=\sum_{j\in S}X_j$. 
For the occurrence of $E_l$, there must be an $S\in {\cal A}$ which differs 
from $S^*$ in an amount $l$ and which has $\| Y-X_S\|^2
\leq \| Y-X_{S^*}\|^2$. 
Let $S$ denote a subset which differs from $S^*$ in an amount $l$.
Here we define $T(S)$ as 
\begin{eqnarray}
T(S)=\frac{1}{2}\left[ \frac{| Y-X_{S} |^2}{\sigma^2}
-\frac{| Y-X_{S^*}| ^2}{\sigma^2} \right],
\nonumber
\end{eqnarray}
where for a vector $x$ of length $n$, 
$|x|^2$ denote $(1/n)\sum_{i=1}^n x_i^2$. 
Then $T(S)\leq 0$ is equivalent to 
$\| Y-X_S\|^2\leq \| Y-X_{S^*}\|^2$. 
The subsets $S$ and $S^*$ have an intersection $S_1=S\cap S^*$ of 
size $L-l$ and a deference $S_2=S\setminus S_1$ of size $l$. 
Note that $X_{S}$ and $Y$ are independent of $X_{S_2}$. 

We use the decomposition $T(S)=\widetilde{T}(S)+T^*$, where 
\begin{eqnarray}
\widetilde{T}(S)=\frac{1}{2}
\left[ \frac{| Y-X_{S} |^2}{\sigma^2}
-\frac{| Y-(1-\alpha)X_{S^*}| ^2}{\sigma^2+\alpha^2P} \right]
\nonumber
\end{eqnarray}
and 
\begin{eqnarray}
T^*=\frac{1}{2}
\left[ \frac{| Y-(1-\alpha)X_{S^*} |^2}{\sigma^2+\alpha^2P}
-\frac{| Y-X_{S^*}| ^2}{\sigma^2} \right].
\nonumber
\end{eqnarray}
For a positive $\tilde{t}=t_{\alpha}$, let $\tilde{E}_l$ denote an event 
that there is an $S\in {\cal A}$ which differs 
from $S^*$ in an amount $l$ and $\widetilde{T}(S)\leq\tilde{t}$. 
Similarly, for a negative $t^*=-t_{\alpha}$, let $E_l^*$ denotes a 
corresponding event that $T^*\leq t^*$. 
Then we have
\begin{eqnarray}
\Pr[E_l] \leq \Pr[E_l^*] + \Pr[\tilde{E}_l].
\nonumber
\end{eqnarray}
First, we evaluate $\Pr[E_l^*]$. 
We use Markov's inequality for $e^{-n\lambda T^*}$ as in \cite{Barron1} with a
parameter $0 \leq \lambda<1/\sqrt{1-\rho_2^2}=1+1/\alpha^2v$.
Then we have 
\begin{eqnarray*}
\Pr[E_l^*]\leq  e^{n\lambda t^*} \mathbb{E}_{Y,X_{S^*}}e^{-n\lambda T^*}.
\end{eqnarray*}
Here we write down the expectation $\mathbb{E}_{Y,X_{S^*}}e^{-n\lambda T^*}$
and apply Lemma \ref{lem_prop} in this paper
as in \cite{IEEE_takeishi}. Then we have for ${\bf x}=(x_1,x_2)^T$
\begin{eqnarray*}
\Pr[E_l^*]\leq e^{n\lambda t^*}
\left(\frac{e^{\phi(L)}}{2\pi}
\int_{-\infty}^{\infty}h_1\sum_{x_1\in{\cal X}_1}
e^{-{\bf x}^T A {\bf x}/2}dx_2\right)^n
\end{eqnarray*}
where $h_1=2/\sqrt{L}$, ${\cal X}_1=\{h_1(k-L/2)| k=0,1,\ldots,L\}$, 
and $A=I-\lambda B$ with the identity matrix $I$ and
\begin{eqnarray}
B=(1-\rho_2^2)\begin{pmatrix} -1 & \frac{1}{\alpha\sqrt{v}} \\
 \frac{1}{\alpha\sqrt{v}} & 1 \end{pmatrix}.
\nonumber
\end{eqnarray}
Then applying Lemma \ref{lem_sm2}, we have
\begin{eqnarray*}
&&\frac{e^{\phi(L)}}{2\pi}
\int_{-\infty}^{\infty}h_1\sum_{x_1\in{\cal X}_1}
e^{-{\bf x}^T A {\bf x}/2}dx_2\\
&\leq&
\frac{e^{\phi(L)}}{2\pi}
\left(1+\frac{\eta A_{11}}{L}\right)
\int_{-\infty}^{\infty}\int_{-\infty}^{\infty}
e^{-{\bf x}^T A {\bf x}/2}dx_1 dx_2\\
&=&
\left(1+\frac{\eta A_{11}}{L}\right)
\frac{e^{\phi(L)}}{\sqrt{ 1-\lambda^2(1-\rho_2^2)}}.
\end{eqnarray*}
Here, using 
\begin{eqnarray*}
A_{11}=1+\lambda (1-\rho_2^2)
\leq 1+\sqrt{1-\rho_2^2}\leq 2,
\end{eqnarray*}
we have 
\begin{eqnarray*}
\frac{e^{\phi(L)}}{2\pi}
\int_{-\infty}^{\infty}h_1\sum_{x_1\in{\cal X}_1}
e^{-{\bf x}^T A {\bf x}/2}dx_2
\leq
\frac{e^{\iota_2}}{\sqrt{ 1-\lambda^2(1-\rho_2^2)}}.
\end{eqnarray*}
Then we have 
\begin{align}
\Pr[E_l^*]&\leq e^{n\lambda t^*}
\left(\frac{e^{\iota_2}}{\sqrt{ 1-\lambda^2(1-\rho_2^2)}}\right)^n\nonumber\\
&=\exp\{-n(\lambda t_{\alpha}+(1/2)\ln(1-\lambda^2(1-\rho_2^2))-\iota_2)\}.\label{el_asta}
\end{align}

Second, we evaluate $\Pr[\tilde{E}_l]$. 
Similarly as the analysis of $E_l^*$, using the parameter $0\leq\lambda\leq 1$, 
we have
\begin{eqnarray}
\Pr[\tilde{E}_l]\!\leq\!\sum_{S_1}\mathbb{E}_{Y,X_{S^*}} 
e^{-n\lambda(\widetilde{T}_1(S_1)-\tilde{t})} 
\!\Bigl(\sum_{S_2} \mathbb{E}_{X_{S_2}} e^{-n\widetilde{T}_2(S)} \Bigr)
^{\lambda},\!\!\!\!\!\label{tilderE}
\end{eqnarray}
where we defined
\begin{eqnarray}
\widetilde{T}_1(S_1)=\frac{1}{2}
\left[ \frac{| Y-X_{S_1} |^2}{\sigma^2+\alpha P}
-\frac{| Y-(1-\alpha)X_{S^*}| ^2}{\sigma^2+\alpha^2P} \right]
\nonumber
\end{eqnarray}
and 
\begin{eqnarray}
\widetilde{T}_2(S)=\frac{1}{2}\left[ \frac{| Y-X_{S} |^2}{\sigma^2}
-\frac{| Y-X_{S_1}| ^2}{\sigma^2+\alpha P} \right].
\nonumber
\end{eqnarray}

As for $\widetilde{T}_2(S)$, 
recalling $C_\alpha=(1/2)\ln(1+\alpha v)$
we can write
\begin{align*}
e^{-n\widetilde{T}_2(S)}
=
\frac{p^h_{Y|X_S}(Y|X_S)}{p^{(c)}_{Y|X_{S_1}}(Y|X_{S_1})}e^{-n C_\alpha },
\end{align*}
where $P^{(c)}_{Y| X_{S_1}}$ is the conditional probability density function
of $Y$ given $X_{S_1}$ in case $X_{ij}\sim N(0,P/L)$,
and 
$p^h_{Y|X_S}$ denotes the conditional probability density function 
$Y$ given $X_{S}$ under the hypothesis that $X_S$ was sent.
Hence we have
\begin{align*}
e^{-n\widetilde{T}_2(S)}
=
\frac{p_{Y|X_{S_1}}(Y|X_{S_1})}{p^{(c)}_{Y|X_{S_1}}(Y|X_{S_1})}
\frac{p^h_{Y|X_S}(Y|X_S)}{p_{Y|X_{S_1}}(Y|X_{S_1})}
e^{-n C_\alpha }.
\end{align*}
Since $(X_{S_1},Y)$ is independent of $X_{S_2}$, we have
\begin{align}\label{tildeS}
\mathbb{E}_{X_{S_2}}e^{-n\widetilde{T}_2(S)}
=
e^{-n C_\alpha }
\frac{p_{Y|X_{S_1}}(Y|X_{S_1})}{p^{(c)}_{Y|X_{S_1}}(Y|X_{S_1})}
\mathbb{E}_{X_{S_2}}
\frac{p^h_{Y|X_S}(Y|X_S)}{p_{Y|X_{S_1}}(Y|X_{S_1})}.
\end{align}
As for the last factor's expectation of (\ref{tildeS}), we have 
\begin{align*}
\mathbb{E}_{X_{S_2}}
\frac{p^h_{Y|X_S}(Y|X_S)}{p_{Y|X_{S_1}}(Y|X_{S_1})}
=
\sum_{X_{S_2}}
\frac{p^h_{Y|X_S}(Y|X_S)p(X_{S_2})}{p_{Y|X_{S_1}}(Y|X_{S_1})},
\end{align*}
where $p(X_{S_2})$ denotes the probability mass function of $X_{S_2}$.
Since $X_S=X_{S_1}\cup X_{S_2}$, and since
$p_{Y|X_{S_1}}(Y|X_{S_1})=p^h_{Y|X_{S_1}}(Y|X_{S_1})$
(because $S_1=S^* \cap S$), 
we have
\begin{align*}
\mathbb{E}_{X_{S_2}}
\frac{p^h_{Y|X_S}(Y|X_S)}{p_{Y|X_{S_1}}(Y|X_{S_1})}
&=
\sum_{X_{S_2}}
\frac{p^h_{Y|X_{S_1}}(Y,X_{S_2}|X_{S_1})}{p^h_{Y|X_{S_1}}(Y|X_{S_1})}\\
&=
\sum_{X_{S_2}}p^h_{X_{S_2}|Y,X_{S_1}}(X_{S_2}|Y,X_{S_1})=1.
\end{align*}
Note that this analysis' idea is same as
that for the corresponding evaluation in \cite{Barron1}.

Hence from (\ref{tildeS}), we have
\begin{align}
\mathbb{E}_{X_{S_2}} e^{-n\widetilde{T}_2(S)}\leq 
 \frac{P_{Y| X_{S_1}}(Y| X_{S_1})}
{P^{(c)}_{Y| X_{S_1}}(Y| X_{S_1})}
 e^{-nC_{\alpha}}.
\label{proddc}
\end{align}
To evaluate the right side of (\ref{proddc}),
we will prove that $P_{Y|X_{S_1}}(Y|X_{S_1})$ 
is nearly bounded by $P^{(c)}_{Y|X_{S_1}}(Y| X_{S_1})$ 
uniformly for all $Y$ and $X_{S_1}$.
Here, we define $Y'=Y-X_{S_1}$ and define $P_{Y'}$ as the probability density
function of each coordinate of $Y'$ and $P_{Y'}^{(c)}$ as $P_{Y'}$ in case 
$X_{ij}\sim N(0,P/L)$.
Then we have
\begin{align}
\frac{P_{Y| X_{S_1}}(Y| X_{S_1})}
{P_{Y| X_{S_1}}^{(c)}(Y| X_{S_1})}
=\prod_{i=1}^n \frac{P_{Y'}(Y'_i)}{P^{(c)}_{Y'}(Y'_i)}.\label{yprime}
\end{align}

Define a set ${\cal X}_2=\{h_2(k-l/2)| k=0,1,\ldots,l\}$
with $h_2=2/\sqrt{l}$.
Note that $Y'=X_{S^*-S_1}+\epsilon$.
Hence, $P_{Y'}(Y_i')$ is the convolution of $N(0,\sigma^2)$ 
and the density of unbiased binomial distribution of size $l$.
Then, by applying Lemma \ref{lem_prop}, we have
\begin{align*}
P_{Y'}(Y'_i)&\leq& 
\frac{e^{\phi(l)}h_2}{2\pi\sqrt{\sigma^2}} 
\! \sum_{w_2\in {\cal X}_2}\!
\exp\left\{-\frac{a_2(w_2-a_3Y'_i)^2+a_4{Y'_i}^2}{2}\right\},
\end{align*}
where $a_2=1+\alpha v$, $a_3=\sqrt{\alpha v/\sigma^2}/a_2$ 
and $a_4=1/(\sigma^2a_2)$ $=(\sigma^2+\alpha P)^{-1}$.

Using Lemma \ref{lem_sm}, we have
\begin{align*}
&h_2
\! \sum_{w_2\in {\cal X}_2}\!
\exp\left\{-\frac{a_2(w_2-a_3Y'_i)^2+a_4{Y'_i}^2}{2}\right\}\\
\leq
& \left(1+\frac{\eta a_2}{l}\right)
\int_{-\infty}^{\infty}
\exp\left\{-\frac{a_2(w_2-a_3Y'_i)^2+a_4{Y'_i}^2}{2}\right\}dw_2.
\end{align*}

Thus, we have
\begin{align}
P_{Y'}(Y'_i)\leq (1+\iota_3) P_{Y'}^{(c)}(Y'_i).\label{dperc}
\end{align}
From (\ref{proddc}), (\ref{yprime}) and (\ref{dperc}), we have
\begin{align}
\mathbb{E}_{X_{S_2}} e^{-n\widetilde{T}_2(S)}\leq 
(1+\iota_3)^n e^{-nC_{\alpha}}.\label{iota3neq}
\end{align}

From (\ref{tilderE}) and (\ref{iota3neq}), we have
\begin{align}
\Pr[\tilde{E}_l]\leq (1+\iota_3)^n
\sum_{S_1}{\mathbb E}_{Y,X_{S^*}}e^{-n\lambda\tilde{T}_1(S_1)}
e^{-n\lambda\Delta_{\alpha}}.
\label{iota_3}
\end{align}
Recall $\Delta_{\alpha}=C_{\alpha}-\alpha R-t_{\alpha}$.

To evaluate the right side of (\ref{iota_3}), 
we will make case argument for (i) $l \leq L - \sqrt{L}$ and 
(ii) $l > L-\sqrt{L}$. 
First, we consider the case (i) $l \leq L - \sqrt{L}$.
In this case, $l'=L-l$ is lager than $\sqrt{L}$. 
According to \cite{IEEE_takeishi}, for 
 $0 \leq \lambda \leq 1$ and ${\bf x}=(x_1,x_2,x_3)^T$, we have
\begin{align*}
& {\mathbb E}_{Y,X_{S^*}}e^{-n\lambda\tilde{T}_1(S_1)}\\
&\leq 
\left(\frac{e^{\phi(l)+\phi(l')}}{(2\pi)^{3/2}}
\int_{-\infty}^{\infty}\!\!h_2h_3\sum_{x_1\in{\cal X}_2}\!
\sum_{x_2\in{\cal X}_3}\!
e^{-{\bf x}^T \tilde{A} {\bf x}/2}dx_3\right)^n,
\end{align*}
where $h_3=2/\sqrt{l'}$, ${\cal X}_3=\{h_3(k'-l'/2)| k'=0,1,\ldots,l'\}$, 
and $\tilde{A}=I-\lambda \tilde{B}$ with the identity matrix $I$ and
\begin{eqnarray}
\tilde{B}=
\left( 
\begin{array}{ccc}
\frac{\alpha v}{1+\alpha v}-
\frac{\alpha^3v}{1+\alpha^2 v}
&
-\frac{ \alpha^2\sqrt{\alpha(1-\alpha)}v}{1+\alpha^2 v}&
\frac{\sqrt{\alpha v}}{1+\alpha v}
-\frac{\alpha\sqrt{\alpha v}}{1+\alpha^2 v}
\\
-\frac{ \alpha^2\sqrt{\alpha(1-\alpha)}v}{1+\alpha^2 v}&
\-\frac{\alpha^2(1-\alpha)v}{1+\alpha^2 v} &
-\frac{ \alpha\sqrt{(1-\alpha)v}}{1+\alpha^2 v}\\
\frac{\sqrt{\alpha v}}{1+\alpha v}
-\frac{\alpha\sqrt{\alpha v}}{1+\alpha^2 v}& 
-\frac{ \alpha\sqrt{(1-\alpha)v}}{1+\alpha^2 v}&
\frac{1}{1+\alpha v} -\frac{1}{1+\alpha^2 v}\\
\end{array} 
\right).
\nonumber
\end{eqnarray}
Applying Lemma \ref{lem_sm3}, we have

\begin{align*}
&\frac{e^{\phi(l)+\phi(l')}}{(2\pi)^{3/2}}
\int_{-\infty}^{\infty}\!\!h_2h_3\sum_{x_1\in{\cal X}_2}\!
\sum_{x_2\in{\cal X}_3}\!
e^{-{\bf x}^T \tilde{A} {\bf x}/2}dx_3\\
&\leq \frac{e^{\phi(l)+\phi(l')}}{(2\pi)^{3/2}}
\left(1+\frac{\eta \tilde{A}_{11}}{l}\right)\left(1+\frac{\eta \tilde{A}_{22}}{L-l}\right)
\int_{\Re^3}
e^{-{\bf x}^T \tilde{A} {\bf x}/2}d{\bf x}\\
&\leq \left(1+\frac{\eta \tilde{A}_{11}}{l}\right)\left(1+\frac{\eta \tilde{A}_{22}}{L-l}\right)
\frac{e^{\phi(l)+\phi(l')}}
{\sqrt{ 1-\lambda^2(1-\rho_1^2)}}\\
&\leq \frac{1+\iota_4}
{\sqrt{ 1-\lambda^2(1-\rho_1^2)}},
\end{align*}
where we used $\tilde{A}_{11}\leq 1$ and $\tilde{A}_{22}\leq 1$.
Thus, we have
\begin{align}
{\mathbb E}_{Y,X_{S^*}}e^{-n\lambda\tilde{T}_1(S_1)}
\leq \left(\frac{1+\iota_4}
{\sqrt{ 1-\lambda^2(1-\rho_1^2)}}\right)^n.\label{iota4}
\end{align}

Now we consider the case (ii) $l > L-\sqrt{L}$.
Since $l'=L-l$ can be small in this case, we cannot use the
same method as the case (i). Instead, we calculate the expectation 
${\mathbb E}_{Y,X_{S^*}}e^{-n\lambda\tilde{T}_1(S_1)}$
specifically, and evaluate the value by using the fact that  $l'$ is small.
The detailed evaluation is written in p.2744r. l-16 -  p.2745l. l-28. of \cite{IEEE_takeishi}.
We have improved the  part of evaluation of applying Lemma 8 in that paper 
by using Lemma \ref{lem_sm} in this paper. 
Namely, the quantity
\begin{align*}
1+\iota_5'=e^{\kappa h_2(B_4+h_2)/\tilde{a}_{33}}
+\frac{\sqrt{\kappa/\tilde{a}_{33}}}{B_4\xi_4'}
e^{-\xi_4'B_4^2/2+\kappa h_2^2/\tilde{a}_{33}}
\end{align*}
is replaced by $1+\eta \tilde{A}_{11}/l$, where $B_4$, $\xi_4'$ and $\kappa$
are defined in \cite{IEEE_takeishi}.
Using $\tilde{A}_{11}\leq 1$, we have
\begin{align}
{\mathbb E}_{Y,X_{S^*}}e^{-n\lambda\tilde{T}_1(S_1)}
\leq \left(\frac{1+\iota_5}
{\sqrt{ 1-\lambda^2(1-\rho_1^2)}}\right)^n.\label{iota5}
\end{align}

From (\ref{iota_3}), (\ref{iota4}) and (\ref{iota5})
\begin{eqnarray*}
\Pr[\tilde{E}_l]
&\leq& 
(1+\iota_3)^n{}_LC_{\alpha L}\frac{(1+\max(\iota_4,\iota_5))^n}
{ (1-\lambda^2(1-\rho_1^2))^{n/2}}e^{-n\lambda\Delta_{\alpha}}\\
&=&{}_LC_{\alpha L}
e^{-n(\lambda\Delta_{\alpha}+(1/2)\ln(1-\lambda^2(1-\rho_1^2))-\iota_1)}
\end{eqnarray*}
where $\iota_1=\ln((1+\iota_3)(1+\max(\iota_4,\iota_5))$. 
Minimizing the right side for $0\leq \lambda\leq 1$, we have 
\begin{eqnarray}
\Pr[\tilde{E}_l]\leq
{}_LC_{\alpha L}
\exp\{-n(D_1(\Delta_{\alpha},1-\rho_1^2)- \iota_1)\}.
\label{el_tilde}
\end{eqnarray}
Thus (\ref{el_asta}) and (\ref{el_tilde}) yield the bound to be obtained.

\section{Proofs of lemmas}
\label{pf_lemmas}
In this section, we prove the lemmas used in Section \ref{sec4}.

\subsection{Proof of Lemma \ref{lem_sm}}
We prove Lemma \ref{lem_sm} by making use of the Euler-Maclaurin formula 
\cite{Bourbaki}, which has several variants.
Among those, we employ the following one
stated as Theorem~1 in \cite{osada1}.
In the statement below,
$b_k$ is the Bernoulli number
($b_0=1$, $b_1=-1/2$, $b_2=1/6$, ...)
and
$B_n(x)$ is the Bernoulli polynomial defined by
\[
B_n(x) = \sum_{k=0}^n {}_n C_k b_{n-k}x^k.
\]
Note that, in \cite{osada1} the residual term is not 
given in the statement but in the proof. 
\begin{thm}[the Euler-MacLaurin formula]\label{EMF}
Let $f(x)$ be a class $C^{2m+2}$ function over $[a,b] \subset \Re$.
Let $\delta = (b-a)/(n+2)$, and $y_k = a+ (k+1)\delta$
($k=-1,0,1,...,n$). Note that $y_{-1}=a$ and $y_{n+1}=b$.
Then
\begin{align} \nonumber
&\delta \Bigl(  \frac{1}{2}f(a) +\sum_{k=0}^{n}f(y_k)+  \frac{1}{2} f(b) \Bigr)
-\int_a^bf(x)dx\\ \label{eq:EMF}
&=
\sum_{j=1}^{m+1}\frac{b_{2j}\delta^{2j}}{(2j)\mbox{\rm !}}
\Bigl(  
f^{(2j-1)}(b)-f^{(2j-1)}(a)
\Bigr)
+R_{m+1}
\end{align}
holds, where
\begin{eqnarray*}
R_{m+1} &=& -\delta^{2m+2}\sum_{k=-1}^{n}J_{k,m+1},  \\
J_{k,m+1} &=& \frac{1}{(2m+2)\mbox{\rm !}}
\int_0^\delta B_{2m+2}\Bigl(\frac{t}{\delta}\Bigr)
f^{(2m+2)}(y_k+t)dt.
\end{eqnarray*}
\end{thm}

For the proof, see \cite{osada1} for example.
We use this theorem with $m=0$, which yields the tightest order result
(Lemma~\ref{lem_sm2})
for our goal. Further, for $m=0$ we can easily obtain some
generalization of the Euler-Maclaurin formula as the following lemma,
with which we can optimize the constant factor of the upper bound in Lemma~\ref{lem_sm2}.

\begin{lem}[some extension of the Euler-Maclaurin formula with $m=0$]\label{EMF2}
For an arbitrary real number $\bar{b}_2$, define $\bar{B}_2(t)=\bar{b}_2-t + t^2$.
Let $f(x)$ be a class $C^{2}$ function over $[a,b] \subset \Re$.
Let $\delta = (b-a)/(n+2)$, and $y_k = a+ (k+1)\delta$
($k=-1,0,1,...,n$). Note that $y_{-1}=a$ and $y_{n+1}=b$.
Then, for all $\bar{b}_2$,
\begin{align} \nonumber
&\delta \Bigl(  \frac{1}{2}f(a) +\sum_{k=0}^{n}f(y_k)+  \frac{1}{2} f(b) \Bigr)
-\int_a^bf(x)dx\\ \label{eq2:EMF} 
&=
\frac{\bar{b}_{2}\delta^{2}}{2}
\Bigl(  
f^{(1)}(b)-f^{(1)}(a)
\Bigr)
-\delta^{2}\sum_{k=-1}^{n}\bar{J}_{k,1}
\end{align}
holds, where
\begin{align*}
\bar{J}_{k,1} = \frac{1}{2}
\int_0^\delta \bar{B}_{2}\Bigl(\frac{t}{\delta}\Bigr)
f^{(2)}(y_k+t)dt.
\end{align*}
\end{lem}

{\it Proof:}
Note that $\bar{B}_2(0)=\bar{B}_2(1)$ 
and $\bar{B}_2'(t)=2B_1(t)=2t-1$ hold.
Using the technique of integration by parts twice, we have
\begin{align*}
\bar{J}_{k,1}=&\frac{\bar{b}_2(f'(y_{k+1})-f'(y_k))}{2}
-\frac{1}{\delta}\int_0^\delta B_1\Bigl( \frac{t}{\delta}\Bigr) f'(y_k +t)dt\\
=&\frac{\bar{b}_2(f'(y_{k+1})-f'(y_k))}{2}
-\frac{f(y_{k+1})+f(y_k)}{2\delta}\\
&+\frac{1}{\delta^2}\int_{x_k}^{y_{k+1}} f(t)dt.
\end{align*}
Summing 
the first side and the third side
of the above from $k=-1$ to $k=n$, we have
\begin{eqnarray*}
\sum_{k=-1}^{n} \bar{J}_{k,1}
\! \!
&=&
\! \!
\frac{\bar{b}_2(f'(y_{n+1})-f'(y_{-1}))}{2}\\
\! \!
&+&
\! \!
\frac{1}{\delta} \Bigl(  \frac{1}{2}f(a) +\sum_{k=0}^{n}f(y_k)+  \frac{1}{2} f(b) \Bigr)
\! \!
+ 
\! \!
\frac{1}{\delta^2}
\! \!
\int_a^b f(t)dt,
\end{eqnarray*}
which yields (\ref{eq2:EMF}).

{\it Remark 1:}
This proof is based on the proof of Theorem~\ref{EMF}
given in \cite{osada1}.

{\it Remark 2:}
In particular with $\bar{b}_2=b_2$,
(\ref{eq2:EMF}) is reduced to the Euler-Maclaurin formula with $m=0$.

{\it Remark 3:}
In the formula in \cite{Bourbaki}, the residual term is given as
\[
-\delta^{2m+3}\sum_{k=-1}^{n}
\frac{1}{(2m+3)\mbox{\rm !}}
\int_0^\delta B_{2m+3}\Bigl(\frac{t}{\delta}\Bigr)
f^{(2m+3)}(y_k+t)dt,
\]
which has 
$f^{(2m+3)}(y_k+t)$ rather than $f^{(2m+2)}(y_k+t)$.
If we use the formula in \cite{Bourbaki}, we have 
\[
I_d/I_c-1=O(s^3),
\]
which is worse order about $s$ than Lemma \ref{lem_sm}.
It is due to higher order derivative of $f(x)$ defined below.
Note that if we make partial integration to the residual term of 
the formula in \cite{Bourbaki}, it yields the same one as 
$R_{m+1}$ in (\ref{eq:EMF}).


Now, we can prove Lemma \ref{lem_sm}.

{\it Proof of Lemma \ref{lem_sm}:}
We define $f(x)=\exp\{-s^2(x-\mu)^2/2\}$. 
Further we define 
\begin{eqnarray*}
I_d'&=& I_d + \frac{f(x_{-1})+f(x_{n+1})}{2}h\\
&=& h\left[\frac{1}{2}f(x_{-1})+\sum_{k=0}^n f(x_k)+\frac{1}{2}f(x_{n+1})\right],\\
I_c'&=& \int_{x_{-1}}^{x_{n+1}}f(x)dx,
\end{eqnarray*}
where we defined $x_{-1}=x_0-h$ and $x_{n+1}=x_n+h$.
Then we have
$
I_d-I_c\leq I_d'-I_c'.
$

Now, we evaluate $I_d'-I_c'$ according to the extended Euler-Maclaurin
formula (\ref{eq2:EMF}), letting
$a=-h(n/2+1)$ and $b=h(n/2+1)$, 
which means
$y_k = x_k$, $\delta = h$, and $2/\sqrt{n}=(b-a)/(n+2)$.
Hence we have
\begin{eqnarray*}
I'_d - I'_c=
h^2\bar{b}_2
\frac{f'(x_{n+1})-f'(x_{-1})}{2}
-h^2 \sum_{k=-1}^{n} \bar{J}_{k,1}.
\end{eqnarray*}

Here, we have
\begin{eqnarray*}
\left|\sum_{k=-1}^{n} \bar{J}_{k,1}\right| \leq \sum_{k=-1}^{n} 
\frac{1}{2}\int_0^h \left| \frac{t^2}{h^2}-\frac{t}{h}+\bar{b}_2
\right|
|f''(x_k +t)|dt.
\end{eqnarray*}
Noting that 
\[
\min_{\bar{b}_2 \in \Re}\max_{0\leq x \leq 1}|x^2-x+\bar{b}_2|=\frac{1}{8},
\]
we can evaluate the above as 
\begin{eqnarray*}
\left|\sum_{k=-1}^{n} \bar{J}_{k,1}\right| 
&\leq&\sum_{k=-1}^{n} \frac{1}{16}\int_0^h |f''(x_k +t)|dt\\
&=&\frac{1}{16}\int_{x_{-1}}^{x_{n+1}} |f''(t)|dt.
\end{eqnarray*}
Thus, we have
\begin{eqnarray*}
I'_d - I'_c \leq \frac{h^2}{16}\left[|f'(x_{n+1})-f'(x_{-1})|
+\int_{x_{-1}}^{x_{n+1}} |f''(t)|dt\right].
\end{eqnarray*}

Now, we evaluate 
\begin{eqnarray*}
f'(x)=-s^2(x-\mu)\exp\left\{-\frac{s^2(x-\mu)^2}{2}\right\}.
\end{eqnarray*}
We can find 
$
\max_{x\in \Re} f'(x)=-\min_{x\in \Re} f'(x)=s/\sqrt{e}
$
and the fluctuation of $f'(x)$ from $x=-\infty$ to $x=\infty$
is $4s/\sqrt{e}$, which is an upper bound on 
$
\int_{x_{-1}}^{x_{n+1}}|f''(t)|dt.
$

Thus, we have
\begin{eqnarray*}
I_d-I_c\leq I_d'-I_c'\leq \frac{h^2}{16}
\left[
\frac{s}{\sqrt{e}}
+\frac{s}{\sqrt{e}}
+\frac{4s}{\sqrt{e}}\right]
=\frac{3sh^2}{8\sqrt{e}}.
\end{eqnarray*}

Recalling $h=2/\sqrt{n}$ and noting $I_c=\sqrt{2\pi}/s$, we have 
\begin{eqnarray*}
I_d\leq \left(1+\frac{3s^2}{2\sqrt{2\pi e}n}\right)I_c=
\left(1+\frac{\eta s^2}{n}\right)I_c,
\end{eqnarray*}
where we have defined $\eta =\sqrt{9/(8\pi e)}$.

\subsection{Proof of Lemma \ref{lem_sm2}}
We prove Lemma \ref{lem_sm2}.
Recall the definition of $I_d$ in this Lemma,
\begin{eqnarray*}
I_d=\int_{-\infty}^{\infty}h\sum_{x_1\in {\cal X}}
\exp\left\{-\frac{{\bf x}^TA{\bf x}}{2}\right\}dx_2.
\end{eqnarray*}

We evaluate the summation about $x_1$ in the above by using Lemma \ref{lem_sm}. 
We can see 
\begin{eqnarray*}
{\bf x}^TA{\bf x}&=&A_{11} x_{1}^2 + (A_{12}+A_{21})x_{1}x_{2}+A_{22} x_{2}^2 \\
&=&A_{11} (x_{1}+a x_2)^2 + b x_{2}^2,
\end{eqnarray*}
where $a$ and $b$ are certain constants depending on $A$.
Thus, we have
\begin{align*}
I_d&=\int_{-\infty}^{\infty}h\sum_{x_1\in {\cal X}}
\exp\left\{-\frac{A_{11} (x_{1}+a x_2)^2+b x_{2}^2}{2}\right\}dx_2 \\
&\leq\left(1+\frac{\eta A_{11}}{n}\right)\!\! \int_{-\infty}^{\infty}\int_{-\infty}^{\infty}
\exp\left\{-\frac{{\bf x}^TA{\bf x}}{2}\right\}dx_1 dx_2\\
&=\left(1+\frac{\eta A_{11}}{n}\right)I_c,
\end{align*}
which proves the lemma.

\subsection{Proof of Lemma \ref{lem_sm3}}
We prove Lemma \ref{lem_sm3}.
Recall the definition of $I_d$ in this Lemma,
\begin{eqnarray*}
I_d=\int_{-\infty}^{\infty}
h_1h_2\sum_{x_1\in {\cal X}_1}\sum_{x_2\in {\cal X}_2}
\exp\left\{-\frac{{\bf x}^TA{\bf x}}{2}\right\}dx_3.
\end{eqnarray*}

We evaluate the summation about $x_1$ and $x_2$ in above by using Lemma \ref{lem_sm}. 
We can see 
\begin{eqnarray*}
{\bf x}^TA{\bf x}=A_{11} (x_{1}+a x_2+b x_3)^2 + c(x_2,x_3),
\end{eqnarray*}
where $a$ and $b$ are certain constants depending on $A$, and $c$ is a quadratic function 
of $x_2$ and $x_3$.
Thus, we have
\begin{eqnarray*}
&&\!\!\!\!h_1\sum_{x_1\in {\cal X}_1}
\exp\left\{-\frac{A_{11} (x_{1}+a x_2+b x_3)^2 + c(x_2,x_3)}{2}\right\} \\
\!\!\!\!&\leq&\!\!\!\!\left(1+\frac{\eta A_{11}}{n}\right)\!\! \int_{-\infty}^{\infty}
\exp\left\{-\frac{{\bf x}^TA{\bf x}}{2}\right\}dx_1.
\end{eqnarray*}

Using the above inequality, we can bound $I_d$ by 
\begin{eqnarray}
\!\!\!\!\!\!\!\!&&\left(1+\frac{\eta A_{11}}{n}\right)\times \nonumber \\
\!\!\!\!\!\!\!\!&&\int_{-\infty}^{\infty}\int_{-\infty}^{\infty}
h_2\!\!\sum_{x_1\in {\cal X}_1}\sum_{x_2\in {\cal X}_2}
\exp\left\{-\frac{{\bf x}^TA{\bf x}}{2}\right\}dx_1 dx_3.
\label{lem6-1}
\end{eqnarray}

With the same way, we can find
\begin{eqnarray*}
{\bf x}^TA{\bf x}=A_{22} (x_{2}+a' x_1+b' x_3)^2 + c'(x_1,x_3),
\end{eqnarray*}
where $a'$ and $b'$ are certain constants depending on $A$, 
and $c'$ is a quadratic function of $x_1$ and $x_3$.
Thus, we have
\begin{eqnarray}
&&\!\!\!\!h_2\sum_{x_2\in {\cal X}_2}
\exp\left\{-\frac{A_{22} (x_{2}+a' x_1+b' x_3)^2 + c'(x_1,x_3)}{2}\right\} \nonumber \\
\!\!\!\!&\leq&\!\!\!\!\left(1+\frac{\eta A_{22}}{n'}\right)\!\! \int_{-\infty}^{\infty}
\exp\left\{-\frac{{\bf x}^TA{\bf x}}{2}\right\}dx_2.
\label{lem6-2}
\end{eqnarray}
According to (\ref{lem6-1}) and (\ref{lem6-2}), $I_d$ is bounded by
\begin{eqnarray*}
\left(1+\frac{\eta A_{11}}{n}\right)\left(1+\frac{\eta A_{22}}{n'}\right)I_c,
\end{eqnarray*}
which proves the lemma.

\section*{Acknowledgment}
The authors thank Professor Andrew R.\ Barron for his valuable comments.
This research was partially supported by JSPS KAKENHI Grant Number JP16K12496.
\ifCLASSOPTIONcaptionsoff
  \newpage
\fi

\end{document}